\documentclass[twocolumn]{aastex631}
\usepackage{appendix}
\usepackage{natbib}
\usepackage{graphicx}
\usepackage{subfigure}
\usepackage{booktabs}
\usepackage{xspace}
\usepackage[encapsulated]{CJK}

\usepackage{amsmath} 

\newcommand{\caii}{\ion{Ca}{2}}
\newcommand{\oi}{\ion{O}{1}}

\newcommand{\feii}{\ion{Fe}{2}}

\newcommand{\msun}{$M_{\odot}$}

\newcommand{\hide}[1]{}

\usepackage[encapsulated]{CJK}

\begin{document}
   
\title{Calcium Triplet Absorption is Common around Little Red Dots}

\author[0000-0002-5612-3427]{Jenny E. Greene}
\affiliation{Department of Astrophysical Sciences, Princeton University, 4 Ivy Lane, Princeton, NJ 08544, USA}

\author[0000-0003-2488-4667]{Hanpu Liu}
\affiliation{Department of Astrophysical Sciences, Princeton University, 4 Ivy Lane, Princeton, NJ 08544, USA}

\author[0000-0002-4684-9005]{Raphael E. Hviding}
\affiliation{Max-Planck-Institut f\"ur Astronomie, K\"onigstuhl 17, D-69117 Heidelberg, Germany}

\author[0000-0003-3729-1684]{Rohan P. Naidu}
\affiliation{Institute for Astronomy, University of Hawai‘i, 2680 Woodlawn Drive, Honolulu, HI 96822, USA}

\author[0000-0003-4075-7393]{David J. Setton}
\thanks{NHFP Hubble Fellow}
\affiliation{The William H. Miller III Department of Physics and Astronomy, The Johns Hopkins University, Baltimore, MD 21218}

\author[0000-0001-9269-5046]{Bingjie Wang (\begin{CJK*}{UTF8}{gbsn}王冰洁\ignorespacesafterend\end{CJK*})}
\thanks{NHFP Hubble Fellow}
\affiliation{Department of Astrophysical Sciences, Princeton University, 4 Ivy Lane, Princeton, NJ 08544, USA}

\author[0000-0002-2057-5376]{Ivo Labbe}
\affiliation{Centre for Astrophysics and Supercomputing, Swinburne University of Technology, Melbourne, VIC 3122, Australia}

\author[0000-0002-5588-9156]{Vasily Kokorev}
\affiliation{Department of Astronomy, The University of Texas at Austin, Austin, TX 78712, USA}

\author[0000-0001-6052-4234]{Xiaojing Lin}
\affiliation{Department of Astronomy, Tsinghua University, Beijing 100084, China}

\author[0000-0003-3596-8794]{Hollis B. Akins}
\affiliation{Department of Astrophysical Sciences, Princeton University, 4 Ivy Lane, Princeton, NJ 08544, USA}
\altaffiliation{NASA Einstein Fellow}

\author[0000-0002-2380-9801]{Anna~de~Graaff}
\affiliation{Max-Planck-Institut f\"ur Astronomie, K\"onigstuhl 17, D-69117 Heidelberg, Germany}

\author[0000-0001-6755-1315]{Joel Leja}
\affiliation{Department of Astronomy \& Astrophysics, The Pennsylvania State University, University Park, PA 16802, USA}
\affiliation{Institute for Computational \& Data Sciences, The Pennsylvania State University, University Park, PA 16802, USA}
\affiliation{Institute for Gravitation and the Cosmos, The Pennsylvania State University, University Park, PA 16802, USA}

\author[0000-0003-2871-127X]{Jorryt Matthee}
\affiliation{Institute of Science and Technology Austria (ISTA), Am Campus 1, 3400 Klosterneuburg, Austria}

\author[0000-0001-5586-6950]{Alberto~Torralba}
\affiliation{Institute of Science and Technology Austria (ISTA), Am Campus 1, 3400 Klosterneuburg, Austria}

\author[0000-0002-3254-9044]{K. Glazebrook}
\affiliation{Centre for Astrophysics and Supercomputing, Swinburne University of Technology, PO Box 218, Hawthorn, VIC 3122, Australia}

\author[0000-0001-5063-8254]{Rachel Bezanson}
\affiliation{Department of Physics and Astronomy and PITT PACC, University of Pittsburgh, Pittsburgh, PA 15260, USA}

\author[0000-0002-3952-8588]{Leindert A. Boogaard} \affiliation{Leiden Observatory, Leiden University, PO Box 9513, NL-2300 RA Leiden, The Netherlands}

\author[0000-0003-2680-005X]{Gabriel Brammer}
\affiliation{Cosmic Dawn Center (DAWN), Copenhagen, Denmark}
\affiliation{Niels Bohr Institute, University of Copenhagen, Jagtvej 128, Copenhagen, Denmark}

\author[0000-0002-0930-6466]{Caitlin M. Casey}
\affiliation{Department of Physics, University of California, Santa Barbara, Santa Barbara, CA 93106, USA}
\affiliation{Cosmic Dawn Center (DAWN), Denmark}

\author[0000-0002-0302-2577]{John Chisholm}
\affiliation{Department of Astronomy, The University of Texas at Austin, Austin, TX 78712, USA}
\affiliation{Cosmic Frontier Center, The University of Texas at Austin, Austin, TX, USA}

\author[0000-0003-3310-0131]{Xiaohui Fan}
\affiliation{Steward Observatory, University of Arizona, 933 North Cherry Ave., Tucson, AZ 85721, USA}

\author{Lukas~J.~Furtak}
\affiliation{Department of Astronomy, The University of Texas at Austin, Austin, TX 78712, USA}
\affiliation{Cosmic Frontier Center, The University of Texas at Austin, Austin, TX, USA}

\author[0000-0001-7201-5066]{Seiji Fujimoto}
\affiliation{David A. Dunlap Department of Astronomy and Astrophysics, University of Toronto, 50 St. George Street, Toronto, Ontario, M5S 3H4, Canada}
\affiliation{Dunlap Institute for Astronomy and Astrophysics, 50 St. George Street, Toronto, Ontario, M5S 3H4, Canada}

\author[0000-0002-2624-3399]{Yan-Fei Jiang (\begin{CJK*}{UTF8}{gbsn}姜燕飞\ignorespacesafterend\end{CJK*})}
\affil{Center for Computational Astrophysics, Flatiron Institute, New York, NY 10010, USA}

\author[0000-0002-0463-9528]{Yilun Ma (\begin{CJK*}{UTF8}{gbsn}马逸伦\ignorespacesafterend\end{CJK*})}
\affiliation{Department of Astrophysical Sciences, Princeton University, 4 Ivy Lane, Princeton, NJ 08544, USA}

\author[0000-0003-2804-0648]{Themiya Nanayakkara}
\affiliation{Sydney Institute for Astronomy, School of Physics A28, The University of Sydney, NSW 2006, Australia.}

\author[0000-0002-7524-374X]{Erica J. Nelson}
\affiliation{Department for Astrophysical \& Planetary Science, University of Colorado, Boulder, CO 80309, USA}

\author{Richard Pan}
\affiliation{Department of Physics and Astronomy, Tufts University, 574 Boston Ave., Medford, MA 02155, USA}

\author[0000-0001-9185-5044]{Eliot Quataert}
\affiliation{Department of Astrophysical Sciences, Princeton University, 4 Ivy Lane, Princeton, NJ 08544, USA}

\author[0000-0003-1282-7454]{Anthony J. Taylor}
\affiliation{Department of Astronomy, The University of Texas at Austin, Austin, TX, USA}
\affiliation{Cosmic Frontier Center, The University of Texas at Austin, Austin, TX, USA}

\author[0000-0003-2919-7495]{Christina C. Williams}
\affiliation{NSF–DOE Vera C. Rubin Observatory/NSF NOIRLab, 950 N. Cherry Ave., Tucson, AZ 85719, USA}
\email{christina.williams@noirlab.edu}

\author[0000-0002-0350-4488]{Adi Zitrin}
\affiliation{Department of Physics, Ben-Gurion University of the Negev, P.O. Box 653, Be'er-Sheva 84105, Israel}

\date{May 2026}

\begin{abstract}
We present new evidence for an optically thick atmosphere surrounding Little Red Dots (LRDs) in the form of Ca absorption at rest-frame 8500~\AA, the calcium triplet (CaT). Building on the detection of CaT absorption in one local LRD  analog (the ``Egg'', \citealt{Lin:2025lowz}), we investigate the region around rest-frame 8500~\AA\ for an archival sample of 17 LRDs ($2 < z < 5$) with \emph{JWST}/NIRSpec grating data. We detect CaT absorption in three individual sources, and find that on average there is an absorption equivalent width (EW) EW$_{\rm CaT} \approx -5$~\AA. Among the three detections, two are approaching the deepest absorption seen in integrated light from stars. Given that the  continuum around CaT is probably dominated by the central engine, we argue that the absorbers are likely to be associated with the LRD. Such absorption has not historically been associated with any components of an AGN central engine, but a cool, optically thick photosphere as proposed for LRDs would naturally produce such absorption. The EW$_{\rm CaT}$ that we observe can be matched by hydrostatic atmosphere models at relatively low metallicity ([M/H]$<-1$) combined with an effective temperature $T_{\rm eff} > 4500$~K. Alternatively, the distribution can be matched at a low photospheric gas density $\rho_{\rm ph}<10^{-11}{\rm~g~cm^{-3}}$ that requires a non-hydrostatic gas structure on dynamical grounds. In the future, metal absorption lines should be a powerful complementary probe of the gas conditions, and possibly the enclosed mass, of LRDs.
\end{abstract}


\section{Introduction}
\label{sec:intro}

\begin{figure*}
     \centering
     \begin{subfigure}
         \centering 
         \includegraphics[width=0.47\textwidth]{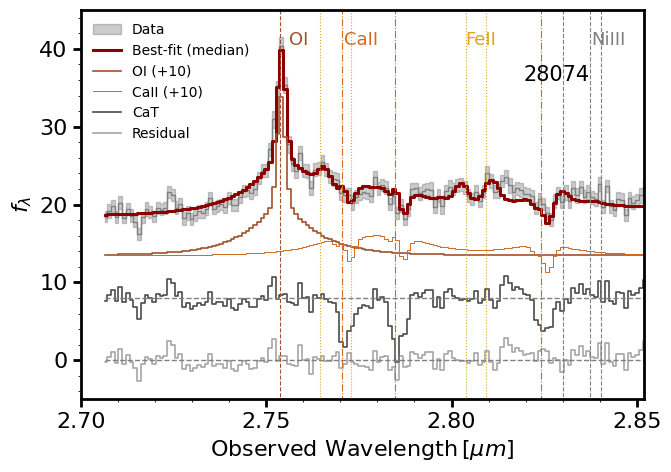}
         \includegraphics[width=0.47\textwidth]{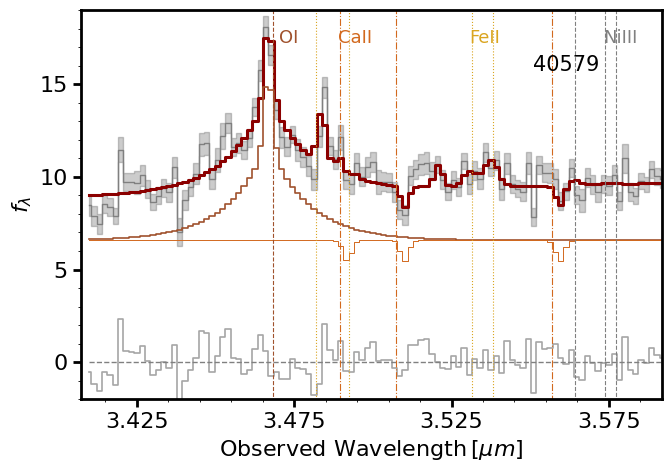}
     \end{subfigure}
     \begin{subfigure}
         \centering
         \includegraphics[width=0.47\textwidth]{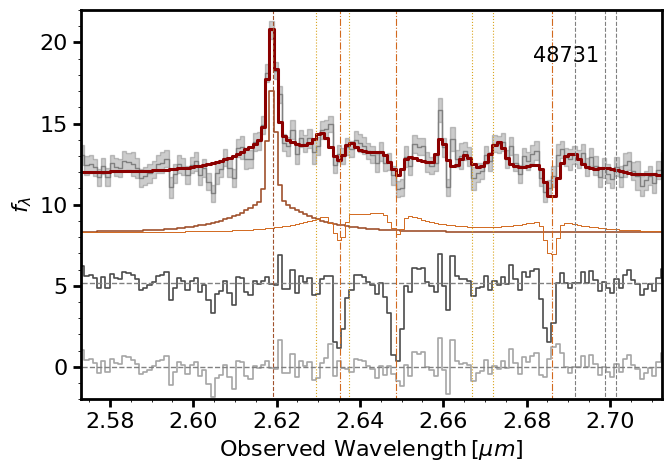}
         \includegraphics[width=0.47\textwidth]{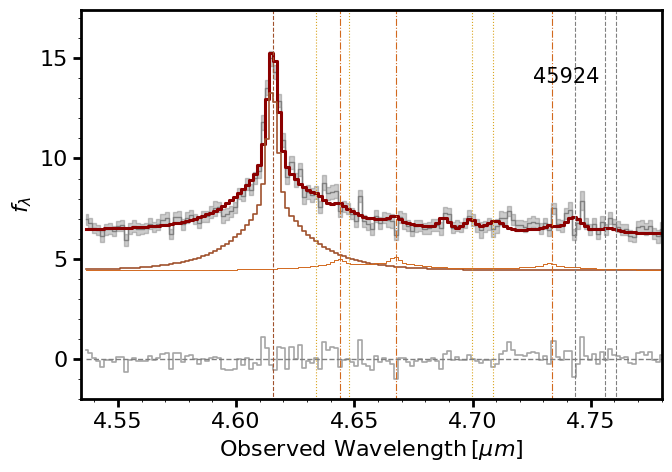}
     \end{subfigure}
     \begin{subfigure}
         \centering
         \includegraphics[width=0.47\textwidth]{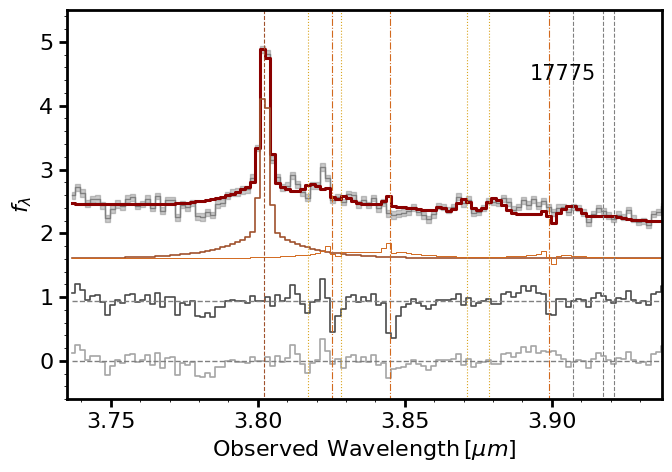}
         \includegraphics[width=0.47\textwidth]{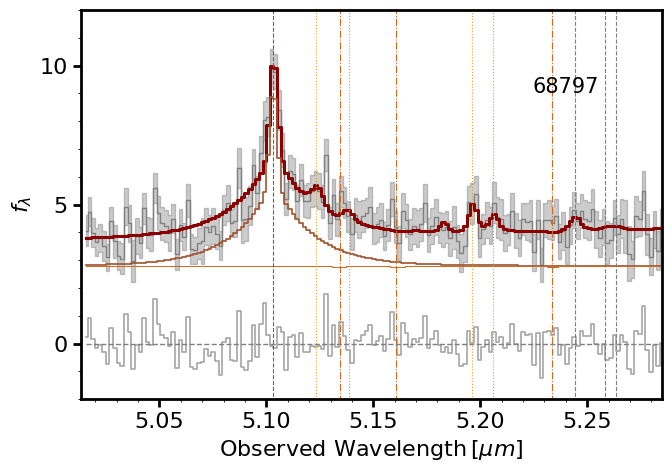}
     \end{subfigure}
        \caption{{\it Left column}: The three sources with detections from the sample. The data are shown in grey, the full model in red. Offset below we show the broad+narrow OI along with the broad+narrow+absorbing CaT. In black, we subtract all models elements except the CaT absorption remaining in black (with a vertical offset), and then the full residual offset below in grey. At top, JADES-28074 ($z=2.27)$ or the `Rosetta Stone' \citep{Juodzbalis:2024} is the most robust example of a target showing evidence for CaT absorption. We also see the clear need for broad CaT emission in this target, as the continuum level to the red of the line is elevated relative to the baseline level blueward of \oi. The GLIMPSE target, ID17775, is a very high S/N spectrum, but the fit is very dependent on the emission model.
        {\it Right column}: Three examples of non-detections. ID40579 \citep[known as the BRD;][]{Wang:2025brd}) is one example with a $3 \, \sigma$ detection, and the residuals clearly show significant structure around the CaT line, although the fit does not prefer CaT in emission.}
        \label{fig:CaTFit}
\end{figure*}

Over the past few years, there has been a blossoming of interest in objects discovered by the \emph{James Webb Space Telescope} that show a qualitatively new spectral energy distribution (SED) from known objects, now known as Little Red Dots \citep{Matthee:2024coin}. They are typically defined by a combination of a `v-shaped' continuum in the rest-frame optical \citep[e.g.,][]{Labbe:2023,Labbe:2023uncover,Barro:2023,Kocevski:2025,Kokorev:2024}, a compact size \citep[e.g.,][]{Furtak:2022,Furtak:2023nature}, and broad Balmer lines \citep[e.g.,][]{Kocevski:2023,Matthee:2024coin,Greene:2024,Lin:2024aspire}. These characteristics are linked, as selecting on any two yields a sample that shares the third characteristic at least 80\% of the time \citep{Hviding:2025}. The v-shaped SED typically has an inflection point at the Balmer limit \citep{Setton:2024break}, and displays a red black-body--like SED redward of this break \citep{DeGraaff:2026dja}, with a range of apparent temperatures that also correlates with the UV properties \citep[e.g.,][]{SunNaidu:2026,Matthee:2026}. 

One clue that the continuum from LRDs must arise from a qualitatively new emission source came from a series of objects with properties that could not be modeled with typical spectral energy distributions. Early on, we struggled to model the strong break in the triply imaged A2744-QSO1 \citep{Ma:2024lens}, and then even more extreme non-stellar breaks were discovered in `The Cliff' \citep{deGraaff:2025} and MoM-BH* \citep{Naidu:2025}. The presence of exponential wings on the broad profiles provided another clue \citep{Rusakov:2025,Chang:2026}. These sources led the community to explore a range of models with high gas density and high covering factor, although the details are still under debate \citep[e.g.,][]{Inayoshi:2025,Liu:2025lrd,Kido:2025, Begelman:2025,Sneppen:2026,MadauMaiolino:2026,ChenY:2026,Chisholm:2026,Nandal:2026}. 

Additional information about the gas conditions surrounding the LRD comes from emission and absorption features rarely seen in typical active galactic nuclei (AGN). Most prominently, Balmer absorption is very common in LRDs \citep[e.g.,][]{Matthee:2024coin,Juodzbalis:2026absorb} but rarely seen in classical AGN \citep[e.g.,][]{Leighly:2025}. \citet{Wangwater:2026} find evidence of water absorption in the near-infrared spectra of a couple of $z \sim 2$ LRDs, which points to the additional presence of molecular gas that is cooler than the apparent $T_{\rm eff} \approx 5000$~K black-body--like continuum seen in the rest-frame optical \citep[e.g.,][]{DeGraaff:2026dja}. Abnormally strong [\ion{Fe}{2}] emission is also seen at least in a subset of LRDs \citep[e.g.,][]{DEugenio:2025,Torralba:2026}.

The calcium triplet (CaT; $\lambda \lambda 8500, 8544, 8665$~\AA) absorption feature offers another powerful opportunity to probe optical depth, pressure, and temperature conditions in the gas surrounding the LRD. As a non-resonant transition, absorption suggests a dense, warm environment to collisionally populate the metastable lower level -- the feature is strong from stellar atmospheres at $T_{\rm eff}\approx5000~$K. One local analog \citep[the Egg;][]{Lin:2025lowz,Ji:2026} shows an EW of $-16.7 \pm 0.5$~\AA\ summed over all lines in the triplet. This EW is deeper than the maximum absorption typically seen in galaxies \citep[EW$_{\rm CaT} \approx -9$~\AA, e.g.,][]{Garcia-Rissmann:2005} and it is deeper than what is seen in stellar population synthesis models \citep{Conroy:2009,Conroy:2010}. Lower surface gravity in the Egg would explain the deeper absorption \citep[e.g.,][]{Lin:2025lowz,Liu:2026}. In fact, even single stars seldom have absorption deeper than the Egg, especially at low metallicity \citep{Cenarro:2001,Cenarro:2002,Lin:2025lowz,Liu:2026}, which emphasizes that we are not seeing a composite stellar population but rather a feature associated with the central engine. 

There is not yet a systematic search for CaT absorption in \emph{JWST-}discovered LRDs at $z>2$. We present a first attempt, as well as quantifying upper limits for the other two SDSS-identified local analogs from Lin et al. Specifically, we compile a sample of grating spectra of LRDs with sufficient signal-to-noise (SN) to enable interesting constraints on the equivalent width (EW) of CaT (\S \ref{sec:data}). In \S \ref{sec:properties}, we characterize the absorbers and argue that they are unlikely to emerge from stellar photospheres in the host, but instead are from the central engine. We confront the measurements with photospheric models in \S \ref{sec:Model}, and present implications and conclusions in \S \ref{sec:Discussion}. Throughout, we assume a concordance cosmology with H$_0$=70, 
$\Omega_{\Lambda} = 0.7$, $\Omega_{M} = 0.3$ \citep{Hinshaw:2013}.

\section{Samples and Data}
\label{sec:data}

In this work, we focus on spectroscopically identified LRDs that also have high signal-to-noise ratio NIRSpec grating spectroscopy from the NIRSpec microshutter array \citep{Ferruit:2022}. To begin, we draw from the comprehensive spectroscopic selection presented in \citet{DeGraaff:2026dja} for a uniform sample of LRDs with NIRSpec/PRISM spectroscopy. The PRISM spectroscopy allows for uniform spectroscopic selection but cannot be used to search for calcium absorption. The selection criteria were tested extensively \citep[][]{Hviding:2025} with the uniformly selected RUBIES sample \citep{Degraaff:2024RUBIES}, and then applied to the full DAWN JWST Archive \citep{Brammer:2025}. The selection criteria are: (a) compactness in the rest-frame optical imaging and (b) v-shaped spectral energy distribution, quantified by UV and optical slopes from the PRISM spectra. As of Fall 2025, this selection yielded 116 objects. 

We then cross-match with the DJA to identify LRDs with relatively deep G235M, G395M or (in one case) G395H grating data. We identify 11 sources with with a redshift range of $2 < z < 5$ and luminosity at $L_{5100\AA} > 3 \times 10^{43}$~erg/s, which is roughly the luminosity above which we can achieve sufficient S/N in grating spectra with moderate 1-5 hour integrations (our one G395H grating observation has a 14 hr integration). We include sources from RUBIES \citep{Degraaff:2024RUBIES}, JADES \citep{Scholtz:2026}, UNCOVER \citep{Bezanson:2022}, and GLIMPSE \citep{Atek:2026, Kokorev:2026}. We note that in the case of GLIMPSE-17775, the source does not have PRISM data and thus is not in the de Graaff et al.\ compilation, although to the extent possible to tell from photometry alone in the UV it would fall in the sample. There is one additional source with only a G235M spectrum, observed by the Deep Dive program, PID 3567/PI Valentino \citep{Ito:2026}. The source was photometrically selected by \citet{Weibel:2026} in a search for sources with strong breaks like the Cliff \citep{deGraaff:2025} but spanning a range of peak wavelength using templates from \citet{SunNaidu:2026}. They searched within extragalactic fields with at least six NIRCam bands \citep{Williams:2025,Weibel:2026}. 

Some of the sources have proprietary data. Our Cycle 4 data (PID 8204, PIs Greene/Labbe) contributes a 3hr NIRSpec/G395M spectrum for A2744-45924, which is also in the de Graaff et al.\ compilation. We have one G395H 5.1 hr integration on another LRD from de Graaff et al.\ (PID 8047 PIs Wang/Nelson). 

We additionally include EMBER LRDs (PID7076, PI Akins) with $z<5.1$. The targets were photometrically selected to be compact and have a v-shaped SED, and then refined based on the PRISM spectra. A very similar set of criteria is applied as in \citet{Hviding:2025,DeGraaff:2026dja}, combining compactness in the NIRCam imaging and PRISM-measured UV and optical slopes. However, targets with slightly redder UV slopes ($\beta_{\rm UV} < 0$) are included. Six EMBER targets pass our threshold ($L_{5100\AA} > 3 \times 10^{43}$~erg/s) to make meaningful CaT measurements. 

In total, we present CaT constraints for (11+6) 17 sources with $2<z<5$ from  \emph{JWST} (Table \ref{tab:Measure}), and two additional new measurements from local analog samples. 

\subsection{Reductions}

Both the archival DJA data and the newer proprietary data are reduced using the latest version 4 reductions of the NIRSpec grating spectra on the DJA\footnote{\protect \url{https://s3.amazonaws.com/msaexp-nirspec/extractions/nirspec_graded_v3.html}},
reduced and processed with the \texttt{msaexp} software \citep{Brammer:2025,Brammer:22}. Data reduction and processing are described in \cite{Degraaff:2024RUBIES, Heintz2025, Valentino2025,Pollock2026}. The EMBER data are reduced with a custom wrapper around the \emph{JWST} pipeline v1.20.2 (H. Akins et al.\ in preparation).

\subsection{CaT EW measurements}

\begin{figure}
\vspace{-1mm}
\hspace{-5mm}
\centering
\includegraphics[width=0.45\textwidth]{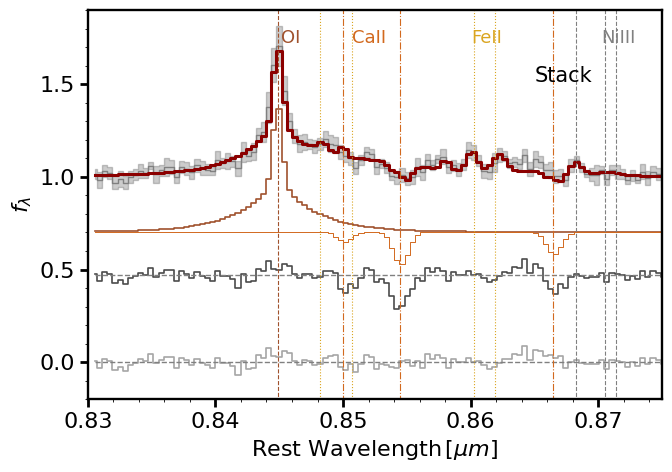}
\caption{
Mean stack of all sources in the CaT region. We marginally detect significant CaT absorption in this stack and the similar version with detections excluded with a similar depth (see Table \ref{tab:Measure}). The full model is shown in black, the data and one-sigma region in grey.
}
\label{fig:Stacks}
\end{figure}

There are a number of relevant spectral features around the CaT (examples in Fig. \ref{fig:CaTFit}). The most prominent emission line is \oi~$\lambda 8449$~\AA. The luminosity of \oi\ is seen to correlate with the H$\alpha$ luminosity in LRDs, suggesting that they arise from a similar region \citep{DeGraaff:2026dja}. We see a wide range of broad to total flux ratios for \oi\ in the LRDs; for instance, the local analogs show no broad \oi\ \citep[e.g.,][]{Lin:2025lowz}, while many of the luminous high-redshift sources show prominent broad 
\oi. Interestingly, we also see evidence for broad calcium triplet emission in a subset of sources. The evidence for the broad Ca in emission is that the overall continuum level is elevated by 10-20\% redward of \oi\ as compared with the blue side. An example of an object with a clear signature of this elevated continuum is shown in MSAID-28074 \citep{Juodzbalis:2024} on the left of Fig.\ \ref{fig:CaTFit}, while on the right we show an example of a source with flat continuum in A2744-45924 \citep{Labbe:2025}. We ascribe this excess to broad Ca emission. 

The fitting is done using the custom code Uniform NIRSpec Inference (Turbo) Engine \citep[{\tt unite};][]{Hviding:2025,hviding_unite}. This code was developed specifically to handle the line-spread function and undersampling of NIRSpec. The empirical line-spread function was derived in \citet{degraaff2024jades}. Because the spectra are undersampled in the spectral direction, the model is integrated within each pixel, rather than evaluated at the pixel center, for data-model comparison. For more details, see \citet{Hviding:2025}. We note that we fit the Magellan+FIRE spectra from \citet{Lin:2025lowz} also using {\tt unite}, and put the line-spread function in by hand in that case. All the fitting procedures are comparable for these two FIRE spectra.

We fit a combination of a narrow+broad Gaussian to the \oi\ line, with a fixed redshift between the two \oi\ components. The line ratios of the CaT lines vary depending on gas conditions. Equal line ratios are seen in the optically thick case \citep[e.g.,][]{Herbig:1980}, which is often assumed for CaT in emission in luminous AGN \citep[e.g.,][]{MartinezAldama:2014}, but these measurements are severely blended with other broad lines. As our fiducial for emission, we take the robustly resolved measurement in the narrow-line Seyfert 1 galaxy 1 Zw I of $0.9:1.0:0.7$ \citep{Rudy:2000}. In typical stars, the absorption line ratios are $\sim 0.3:1.0:0.7$ \citep{Jones:1984}, and we take these ratios as our fiducial for the absorption. However, in the Egg, the absorption line ratios are observed to be $0.7:1.0:0.7$. To explore the systematic effects of these ratios on the measured EWs, we test the impact of swapping the fiducial values with $1:1:1$ for both the absorption and the emission. 

Motivated by classical AGN studies, where the line shape of \oi\ and CaT broad emission are consistent with each other \citep[e.g.,][]{Persson:1988}, we fix the line shape of \caii\ emission to the \oi\ model. Because both transitions (and the \feii) are relatively low ionization lines, they are found to arise from a similar region of the broad-line region in models \citep[e.g.,][]{Joly:1989,Panda:2020} as well. This is a strong, but well-justified prior, and our fits are not stable without it. We also set a prior of FWHM$\leq 500$~km/s on the CaT absorption line width. Without this prior, the CaT absorption and broad emission can each grow to very large values and cancel each other out. We caution that the fitted linewidth and velocity offset are both quite sensitive to the line-width prior. For instance, if we relax the prior on JADES-28074 to FWHM$\leq 1500$~km/s, both FWHM and velocity offset change by a factor of two. Higher-resolution spectra are needed to relax these priors. 

Since different sources clearly show either a significant additional source of emission redward of the \oi\ line (Fig.\ \ref{fig:CaTFit} left) or very little (Fig.\ \ref{fig:CaTFit} right), we build two models, with and without \caii\ in emission. We explore adding higher-order Pa lines but do not find an improvement to our fits\footnote{We do not have uniform coverage of brighter Pa lines in the data-set to constrain these weak lines.}. We choose the more appropriate fit for each object based on the Widely Applicable Information Criterion (WAIC). If the WAIC with the \caii\ emission included is more negative by at least 11.8, roughly corresponding to a $2 \sigma$ improvement \citep[e.g.,][]{Hviding:2025}, then we adopt the fit including \caii\ in emission as the fiducial fit. 

We must also include additional narrow lines in the fit. As has been emphasized in many papers \citep[e.g.,][]{DEugenio:2025,Torralba:2026,Lin:2025lowz,Ji:2026}, the narrow forbidden [\feii] emission is very strong in LRDs, and there are a number of transitions in this region that must be modeled ([\feii]~$\lambda \lambda \lambda 8482, 8507, 8603, 8619$~\AA). We also include \ion{Cl}{2}~$\lambda 8581$~\AA, and NI$\lambda \lambda \lambda 8683,8706,8714$~\AA, with the latter tied to a flux ratio of 1:0.16:0.19. The resulting CaT EW and broad \oi\ properties are summarized in Table \ref{tab:Measure}. Throughout, when we refer to EW$_{\rm CaT}$, we mean the sum of the three lines, and only the absorption component. The Egg measurement was made in a very comparable fashion with single Gaussian absorbers. We note that with the current data, and given the challenges of robustly modeling the broad CaT emission, we cannot say for sure whether the absorber sits in front of the broad-line region, as is clearly the case for the Balmer absorbers \citep[e.g.,][]{DEugenio:2025,Matthee:2026,Ji:2025}.

We determine that a CaT measurement is significant if the depth is more than $5 \, \sigma$ from zero. While we measure the P16 and P84 values from the MCMC fit, we find that the uncertainties are very symmetric in practice (Table \ref{tab:Measure}). In total, we have significant detections in three of the 17 objects (Table \ref{tab:Measure}). There are two other objects (ID40579 and ID154183) with detections at the $3 \, \sigma$ level that we view as less reliable.

\subsection{Stacking}

\begin{figure}
\vspace{-1mm}
\centering
\includegraphics[width=0.45\textwidth]{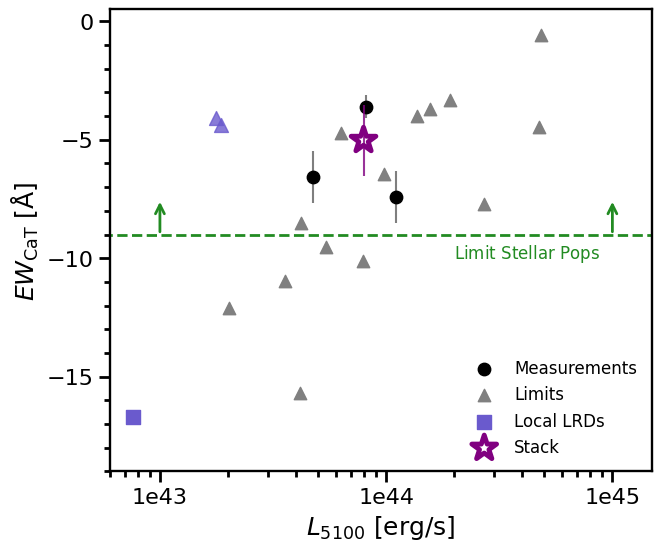}
\caption{
The CaT equivalent width measurements and limits plotted against luminosity $L_{5100}$. Limits are shown as upward-facing triangles. The two blue triangles are the other low-redshift LRDs from \citet{Lin:2025lowz}. The stacked detection is shown as a star. We considered a relationship between EW$_{\rm CaT}$ and $L_{5100}$ (i.e., optical luminosity), because the strongest source (the Egg) is at considerably lower $L$. However, with this limited sample we do not find statistically significant evidence of correlation. 
}
\label{fig:LCaT}
\end{figure}

We do a very simple stacking of the G395M grating data in our sample, including the archival DJA and EMBER objects but excluding the one G395H object. We stack both with and without the detections included, to see whether there is evidence of CaT absorption in the population even when individual detections are excluded. In addition to stacking all non-detections, we also look for trends in stacks of luminosity, UV slope, and Balmer break strength, but the statistics are very limited and no clear trends emerge.

We place the systemic redshift at the velocity of the \oi\ line, as the strongest emission feature that we have in common across all spectra. We test whether the \oi\ is a good tracer of the systemic velocity, and verify that the redshift traced by \oi\ is consistent (within 30~km/s) with at least one other strong and isolated narrow line (although we use the narrow Pa$\beta$ for RUBIES-40579 and VALENTINO-48731). 
Each spectrum is normalized to the median flux within the $0.88-0.90$~\micron\ rest-frame wavelength range, to avoid the CaT region itself. We interpolate each spectrum onto the same wavelength grid, taking a wavelength sampling set by the median number of pixels across the sample. We then take the mean value at each wavelength in the normalized spectra. We test the ramifications of doubling the wavelength sampling or taking the median rather than the mean, and find the median to be noisier. Note that the line spread function is a function of wavelength, which thus varies at the position of the CaT absorber for different objects in the stack.

To measure absorption in the stack, we repeat the same fitting procedure as for the single objects, attempting a fit with and without CaT emission and picking the emission only if it is preferred by the fit. We also bootstrap 100 additional stacks, with replacement. We fit these with the preferred model to determine the uncertainties on the stacked measurements. The mean stacked spectrum of the 16 sources (WANG-28812 excluded) is shown in Figure \ref{fig:Stacks}. We detect $3 \, \sigma$-level absorption in the stack (Table \ref{tab:Measure}). We have only a $\sim 2 \, \sigma$ hint of CaT absorption when the detections are excluded.

\begin{figure}
\vspace{-1mm}
\hskip -5mm
\centering
\includegraphics[width=0.45\textwidth]{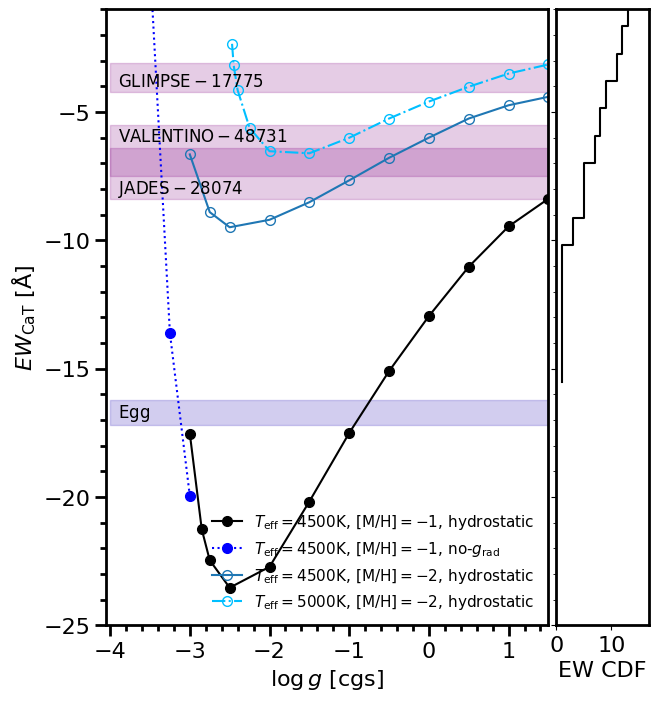}
\caption{
Predicted CaT EW from the synthetic spectral library in \citet{Liu:2026}. The surface gravity parameter $\log g$ is a proxy for photospheric gas density. On the right, we show the distribution of CaT EW limits from the data, to facilitate comparison. 
The hydrostatic models at $\rm[M/H]=-1$ agree with the Egg at $\log g\sim-3$ or $-1$ but overpredict the EW for the new measurements presented here. Lowering $\rm[M/H]$ and raising $T_{\rm eff}$, or alternatively reducing $\log g$ and landing on the super-Eddington (``no-$g_{\rm rad}$'') branch, yields lower EW. As a caveat, measurement systematics can make direct comparison challenging. The biggest single effect is in the unknown line ratios (e.g., JADES-28074 could have EW$_{\rm CaT} \approx -12$~\AA). Other systematics (line or continuum infill) would tend to make the measured lines weaker than the models.}
\label{fig:HANPU}
\end{figure}

\subsubsection{Systematic uncertainties}

Given the spectral resolution in G395M, along with the spectral complexity of modeling CaT in both emission and absorption, we do face some significant degeneracy in our fits. To limit degeneracy,  we fix the CaT emission-line shape to that of \oi. We also fix the relative depths of the CaT emission and absorption. In our tests, the encouraging thing is that our three sources remain as $5 \, \sigma$ detections in all permutations of absorptin and emission-line ratios. Also, varying the emission line ratios between the 1~Zw~I value and the optically thick case does not change our result by more than $\sim 0.5$~\AA. However, if we shift the absorption to the optically thick case (1:1:1), then we can see up to 5~\AA\ deeper absorption -- translating to EW$_{\rm CaT} \approx -12$~\AA\ in JADES-28074 (the Rosetta Stone). We see no preference in the fits in either $\chi^2$ or WAIC between the two line ratios, and so we identify this systematic uncertainty in the CaT absorption line ratio as our biggest systematic.

We also note that we limit the line width of the CaT absorption to FWHM$=500$~km/s, to mitigate  degeneracy between broad CaT emission and absorption if we allow the line-width to be free. In other words, we do not have a strong constraint on a possible broader line component. Finally, there is also the possibility of low levels of infill from high-order Paschen lines. Given that we do not see strong emission infill in the stacks, infill is quite subdominant to the line ratio uncertainty. Understanding the real distribution of strengths across the population will require deeper and higher-resolution data.

\section{Properties of CaT Absorption}
\label{sec:properties}

The first important outcome of this paper is that, on average, there is EW$_{\rm CaT}\approx -5$~\AA\ in \emph{JWST-}discovered LRDs (Figure \ref{fig:Stacks}), so CaT absorption appears to be a frequent occurrence in LRDs. We now turn to the question of the origin of the absorption, by investigating the fitted properties. While our arguments are not definitive, we prefer a scenario in which the absorption emerges within the gas around the LRD and not from stellar photospheres. 

\subsection{Line Properties}

We detect CaT absorption individually in only three sources, giving us limited ability to measure line properties. However, in addition to the EW of the absorber, we also measure a line width and a velocity relative to the \oi\ emission line (Table \ref{tab:linewidth}). Given that the likely stellar mass of LRD hosts is $M_* \sim 10^8$~\msun\ based on clustering \citep[e.g.,][]{Lin:2024aspire,Matthee:2024}, the dispersions that we observe are much higher than what one might expect. However, an optically thick CaT line can have a broad emergent profile due to saturation in its wings, so its FWHM need not directly trace the gas velocity dispersion \citep[e.g.,][ in the case of the Egg]{Liu:2026}. Thus the linewidth alone is inconclusive. The relative velocities do not teach us much more. Only in the case of JADES-28074 do we see even a small hint for a redshift relative to systemic. In the future, similar measurements for a larger sample would be interesting to compare with the properties of the Balmer absorption.

\subsection{Correlations}

With mostly upper limits, it is hard to evaluate whether the presence of CaT absorption is correlated with other properties. Because the Egg has the deepest CaT absorption and the lowest luminosity in the sample, we looked for evidence of a trend between $L_{5100}$ and CaT EW (Figure \ref{fig:LCaT}, left) and find no statistically significant evidence of correlation. We also note that the strongest absorbers are detected at the lower redshift end of the sample. In principle this could point towards changing conditions (e.g., metallicity) with redshift. Alternatively, these sources are brighter, so we cannot rule out that this is in part a signal-to-noise effect.

\subsection{Nonstellar CaT Absorption}

Now we try to evaluate the likely origin of the CaT absorption. In the case of the Egg, the absorption is so strong that it is inconsistent with stellar populations \citep{Lin:2025lowz}. In the wider sample, we cannot rule out a contribution from stellar absorption from the EWs alone, nor do the line properties give much additional information. On the one hand, the detections and the average absorption are both at levels that are seen in integrated stellar populations, with a limiting depth EW$_{\rm CaT}\approx -9$~\AA\ \citep{Garcia-Rissmann:2005}. On the other hand, if the CaT absorption comes from the stellar population, then we can estimate the maximum fraction of the light at 0.8$\, \micron$ that could come from the non-stellar component, i.e. the central engine. Specifically, even if we assume that the stellar absorption is maximally deep at $-9$~\AA, the fact that we observe $\sim -5$~\AA\ of absorption means that the starlight must contribute at least 50\% of the continuum light (or more if the lines are intrinsically weaker). 

There are many reasons to doubt that star-light contributes $>50\%$ of the 0.8~$\micron$ continuum. One is the compact sizes. In the UV, a range of morphologies are observed \citep[e.g.,][]{Killi:2024,Chen:2025host, SunNaidu:2026,Cloonan:2026,Baggen:2026,Golubchik:2025}, but beyond 0.4$\, \micron$, we observe a uniformly compact morphology \citep[e.g.,][]{Furtak:2023nature,Baggen:2024,Golubchik:2025,Yanagisawa:2026}. Modeling this continuum as starlight-dominated leads to very high stellar masses \citep[e.g.,][]{Labbe:2023} and stellar densities \citep[e.g.,][]{Ma:2024lens,Baggen:2024}. Finally, we also observe Balmer breaks that cannot easily be modeled with stellar populations \citep{Ma:2024lens,deGraaff:2025,Naidu:2025,Matthee:2025env,Lin:2024aspire,Setton:2024break,Setton:2025}. On net, then, we still consider the most likely origin of the CaT absorption to be from the central engine and not stars in the host galaxy.

\begin{deluxetable*}{lclrccc}
\tablehead{
\colhead{Name} & \colhead{$z$} & \colhead{Ca} & \colhead{EW$_{\rm CaT}$}  & \colhead{log ($L_{5100}$)} & \colhead{$\beta_{\rm UV}$} & \colhead{Balmer break} \\
\colhead{(1)} & \colhead{(2)} & \colhead{(3)} & \colhead{(4)} & 
\colhead{(5)} & \colhead{(6)} & \colhead{(7)} }
\startdata
JADES-12402 & 3.191 & 0 & $-2.14^{+1.39}_{-2.15}$ & 43.62 & $-0.619$ & 2.61 \\
RUBIES-144195 & 3.355 & 0 & $-2.50^{+1.85}_{-3.03}$ & 43.55 & $-1.45$ & 3.28 \\
RUBIES-154183 & 3.546 & 0 & $-5.35^{+1.56}_{-1.69}$ & 43.9 & $-0.707$ & 9.41 \\
GLIMPSE-17775 & 3.5 & 1 & $-3.60^{+0.58}_{-0.49}$ & 43.91 & \nodata & \nodata \\
JADES-28074 & 2.259 & 1 & $-7.42^{+1.08}_{-1.08}$ & 44.04 & $-1.25$ & 1.57 \\
WANG-28812 & 4.222 & 0 & $-5.21^{+3.32}_{-3.80}$ & 43.62 & $-0.531$ & 1.72 \\
RUBIES-40579 & 3.105 & 0 & $-2.21^{+0.61}_{-0.66}$ & 44.14 & $-0.365$ & 3.01 \\
A2744-45924 & 4.463 & 0 & $-1.79^{+0.85}_{-0.85}$ & 44.68 & $-0.602$ & 3.17 \\
VALENTINO-48731 & 2.100 & 1 & $-6.56^{+1.10}_{-1.09}$ & 43.68 & \nodata & \nodata \\
JADES-68797 & 5.040 & 0 & $-0.65^{+0.44}_{-0.83}$ & 44.19 & $-1.11$ & 2.16 \\
JADES-73488 & 4.132 & 0 & $-2.42^{+1.70}_{-3.08}$ & 43.31 & $-1.71$ & 1.14 \\
\hline
EMBER-180348 & 4.374 & 0 & $-1.59^{+0.96}_{-1.08}$ & 43.8 & $-1.69$ & 1.23 \\
EMBER-249300 & 3.01 & 0 & $-2.93^{+1.27}_{-1.26}$ & 43.99 & $-1.14$ & 1.02 \\
EMBER-541981 & 2.956 & 0 & $-0.58^{+0.43}_{-0.77}$ & 44.28 & $-1.7$ & 1.45 \\
EMBER-752314 & 4.526 & 0 & $-2.00^{+1.31}_{-1.80}$ & 44.43 & $-1.33$ & 3.4 \\
EMBER-788483 & 4.016 & 0 & $-0.09^{+0.07}_{-0.13}$ & 44.68 & 0.809 & 1.93 \\
EMBER-9743 & 3.975 & 0 & $-4.67^{+1.73}_{-1.72}$ & 43.74 & $-0.368$ & 1.49 \\
\hline
STACK-ALL(16) & \nodata & 1 & $-5.0 \pm 1.4$ & 43.9 & \nodata & \nodata\\
STACK-NoDet(13) & \nodata & 0 & $-4.3 \pm 1.8$ & 44.0 & \nodata & \nodata \\
\enddata
\label{tab:Measure}
\caption{Column 1: Survey and MSAID. Column 2: Redshift measured from the \oi\ line. Column 3: Flag for significant CaT absorption detection. Column 4: EW$_{\rm CaT}$ and $16$ and 84\% values from the summed posterior of all three transitions. Column 5: Log optical luminosity at 5100\AA\ (erg/s). Column 6: UV slope $\beta_{\rm UV}$ measured from PRISM spectra. Column 8: Balmer break strength following \citet{DeGraaff:2026dja}.}
\end{deluxetable*}

\section{Model Comparison}
\label{sec:Model}

A number of LRD models have emerged to explain the intrinsically cool continua of LRD spectra along with their prominent Balmer breaks and Balmer absorption. These models have invoked a dense gas envelope or cocoon \citep[e.g.,][]{Inayoshi:2025,Begelman:2025,Liu:2025lrd,Sneppen:2026}. The fact that we see CaT in absorption, which is unlikely to be from normal stars in the galaxies, provides additional motivation for the idea of an optically thick photosphere. Here we explore attributing the absorption to such a photosphere and investigate whether more might be said about the gas conditions in this particular scenario. We specifically compare with the grid of optically thick, plane-parallel, thermalized atmospheres that reproduce the red colors and black-body--like continua of the LRDs from \citet{Liu:2026}. The Liu et al.\ models are able to simultaneously match the continuum colors and CaT absorption EW of the Egg for an effective temperature $T_{\rm eff}=4500{\rm~K}$ and metallicity $[\rm M/H]=-1$ with a photospheric density $\rho_{\rm ph}\sim10^{-11}{\rm~g~cm^{-3}}$ (corresponding to the surface gravity parameter $\log g\sim-3$, Figure~\ref{fig:HANPU}); an alternative solution at $\rho_{\rm ph}\sim10^{-9}{\rm~g~cm^{-3}}$ or $\log g\sim-1$ would give a similar EW but a worse match in the IR continuum. We note that we are assuming solar-scaled abundance ratios in our calculations, which also could in principle impact the Ca depth both directly (by changing the Ca line opacity) and indirectly (by affecting the thermal and ionization balance).

Here, we ask whether the same model grid can explain the wide range of observed CaT depths presented in this paper, for a reasonable range in density, effective temperature, and metallicity. We first note that while scattering \citep[e.g.,][]{Laor:2006,Rusakov:2025,Torralba:2026,Matthee:2026} could broaden the CaT absorption line and make the apparent depth lower, we do not think that scattering alone can drive the difference in depth across the sample. For one thing, the Egg itself shows very strong scattering wings (de Graaff, A.\ in preparation). 

In Figure \ref{fig:HANPU}, we compare the fiducial model of the Egg with the distribution of EW measurements and limits seen in the data. As in \citet{Liu:2026}, we interpret $\log g$ mainly as a proxy for the photospheric density, which the radiative transfer depends on directly\footnote{The surface gravity would naively suggest a dynamical mass within the photosphere $M_{\rm dyn}=2\times10^3~M_\odot~L_{44}g_{-3}T_{5000}^{-4}$, where $L_{\rm 44}\equiv L/(10^{44}{\rm~erg~s^{-1}})$, $g_{-3}\equiv g/(10^{-3}{\rm~cm~s^{-2}})$, $T_{\rm 5000}\equiv T_{\rm eff}/(5000{\rm~K})$, but the true mass could differ significantly in the presence of turbulent or bulk motion \citep[see discussion in \S 2 of ][]{Liu:2026}.}.
The optically thick atmospheres in \citet{Liu:2026} cover a density range $10^{-11}{\rm~g~cm^{-3}}\lesssim\rho_{\rm ph}\lesssim10^{-7}{\rm~g~cm^{-3}}$ or $-3\lesssim \log g\leq1.5$ in their dynamically self-consistent (``hydrostatic'') calculations. The lower bound of the range (which depends on $T_{\rm eff}$) is reached where radiative acceleration overcomes gravity throughout the atmosphere and makes a hydrostatic solution impossible. We disfavor the high end $\log g \gg 1.5$ ($\rho_{\rm ph} \gg 10^{-8}{\rm~g~cm^{-3}}$) for most LRDs because, under hydrostatic equilibrium, this would imply $M_{\rm BH}\gg10^8~M_\odot$, assuming a typical luminosity of $10^{44}{\rm~erg~s^{-1}}$ and $T_{\rm eff}=5000{\rm~K}$. This mass is higher than that suggested by standard virial relations \citep[e.g.,][]{Juodzbalis:2026} for most LRDs, and various recent arguments favor much lower masses \citep[e.g.,][]{Rusakov:2025,Greene:2026,Naidu:2026,Sun:2026mass}.

The range in observed EW could be attributed to a range in metallicity. For the fiducial tracks, we chose a metallicity of [M/H] $=-1$ because it seems most plausible given the likely mass scale of the hosts. If we take the clustering measurements at face value \citep[e.g.,][]{Matthee:2025env,Lin:2025}, or estimate from host-LRD decomposition \citep{SunNaidu:2026}, then the stellar mass range of $M_* \approx 10^8 \, M_{\odot}$ would yield [M/H] $\approx -1$ \citep[e.g.,][]{Lewis:2026}. As Figure \ref{fig:HANPU} shows, at the default $T_{\rm eff} = 4500$~K and [M/H] $\approx -1$, the hydrostatic models have stronger CaT absorption than the majority of the data over our full range of plausible densities. Lowered metallicity reduces the predicted absorption strength toward better agreement with the data. However, even assuming a lower value of $\rm[M/H]=-2$ and solar abundance ratios, the model atmosphere at conditions inferred for the Egg still produces EW = $-6.6$~\AA, which is ruled out for seven of the sources measured in this work. We also tested increasing $T_{\rm eff}$ to 7500~K while keeping [M/H]$=-1$, which predicts an equivalent width EW$_{\rm CaT} < -4$~\AA, stronger than at least three limits presented here. 

On the other hand, if we simultaneously consider a lower metallicity of $\rm[M/H]=-2$ and a higher effective temperature of $T_{\rm eff} = 5000$~K (cyan line in Fig.\ \ref{fig:HANPU}), then we can naturally reach the full range of measurements and limits presented here.  Therefore, one possibility to describe the difference between the Egg and the sources studied in this work is that a combination of higher effective temperatures \citep[e.g.,][]{deGraaff:2026} and lower metallicity explains the shallower CaT absorption. Even lower metallicity, $\rm[M/H]<-2$, could also explain the level of absorption, and may characterize at least some LRDs \citep[e.g.,][]{Nikopoulos:2026}. If indeed EW$_{\rm CaT}$ is growing deeper at lower redshift, then perhaps it could prove a proxy for metal-enrichment in these sources. 

Density differences could also be at play. The CaT EW of the $T_{\rm eff}\sim5000{\rm~K}$ models in \citet{Liu:2026} non-monotonically depends on $\log g$, being deepest at $\log g\sim-2.5$ ($\rho_{\rm ph}\sim$ a few$\times10^{-11}{\rm~g~cm^{-3}}$) and weakening towards both ends. Therefore, if we assume that the effective temperature and metallicity inferred for the Egg also apply to other LRDs, lowered photospheric densities outside the nominal range could also produce shallower absorption. Specifically, there is a log $g < -3$ branch in the Liu et al.\ models (labeled ``no-$g_{\rm rad}$'' in Figure \ref{fig:HANPU}) that also predicts lower CaT at lower $\rho_{\rm ph}$. Apart from higher $T_{\rm eff}$ combined with low $\rm[M/H]$, the second feasible explanation within the model framework, is that the majority of the LRDs occupy the parameter space $\rho_{\rm ph}<10^{-11}{\rm~g~cm^{-3}}$. No dynamically self-consistent hydrostatic solution exists at such low $\log g$, due to radiative acceleration exceeding gravity (i.e., super-Eddington) at all depths. If LRDs are mostly in that density regime, they cannot have hydrostatic atmospheres. 

We caution that the $\log g<-3$ branch in \citet{Liu:2026} (but not their higher-$\log g$ models) ignores the above-mentioned radiative acceleration and therefore is not dynamically self-consistent; hence the label ``no-$g_{\rm rad}$'' in Figure~\ref{fig:HANPU}. In addition to the dynamical concerns, at such low $\log g$, the atmosphere conditions likely start to deviate from the plane-parallel and LTE assumptions in the models. The quantitative predictions of the EW are therefore unreliable. However, the higher-density, hydrostatic models lie in the regime where the model assumptions are valid, so the overprediction is more robust. 
Indeed, the plane-parallel geometry is accurate for a hydrostatic atmosphere at $\log g\gg-4$ due to the large inferred photosphere radius of LRDs \citep[Equation 5]{Liu:2026}. As for uncertainties due to non-LTE atomic populations, past studies in the stellar atmosphere regime have found only weak non-LTE corrections \citep{Jorgensen:1992,Mashonkina:2007,Merle:2011}, and so did the test in Liu et al.\ at $\rho_{\rm ph} \sim 10^{-8.5}$ or $10^{-10.5}{\rm~g~cm^{-3}}$. Moreover, these corrections are in the direction of enhancing the predicted EW, which would exacerbate the model-data discrepancy. 

In summary, the sheer detection of CaT absorption in individual objects and stacks points towards an envelope or cocoon that is optically thick in the CaT line, which naturally produces absorption from stratification in the excitation levels of \ion{Ca}{2}. Within the framework of atmosphere models, the fact that the EWs are typically shallow compared to the Egg either points to simultaneously higher temperature and lower metallicity, or to gas conditions with $\rho_{\rm ph} < 10^{-11}{\rm~g~cm^{-3}}$, and dynamical (not hydrostatic) conditions. Future joint modeling of SEDs and absorption features is the clear next step (\S \ref{sec:Discussion}).

Furthermore, we cannot discount alternative scenarios outside our model framework. For example, the model grid in \citet{Liu:2026} is based on hydrostatic atmospheres, whereas LRDs may feature a pseudo-photosphere in a dynamic gas layer, analogous to stellar winds or explosions \citep{Martins:2026,Chisholm:2026,Naidu:2026}. We discuss possible ramifications of different dynamical configurations in \S \ref{sec:Discussion}. Alternatively, if the continuum-emitting gas layer is not fully thermalized at its bottom but is instead illuminated by some ionizing radiation hotter than the local temperature, as in the cocoon scenario \citep[e.g.,][]{Naidu:2025,Sneppen:2026}, the atomic level population could deviate from LTE throughout the medium and change the predictions, perhaps even ionizing all of the \ion{Ca}{2}.

\begin{deluxetable}{lcc}
\tablehead{\colhead{NAME} & \colhead{FWHM$_{\rm CaT}$ (km/s)} & \colhead{$\Delta v$ (km/s)}}
\startdata
GLIMPSE-17775 & $100\pm40$ & $26\pm9$ \\
JADES-28074 & $370\pm40$ & $70\pm20$ \\
VALENTINO-48731 & $230\pm40$ & $14\pm13$ \\
\enddata
\label{tab:FWHM}
\caption{The three sources with CaT absorption. We tabulate the absorption line-width (km/s) and the velocity offset between the \oi\ narrow velocity and the absorber.}
\label{tab:linewidth}
\end{deluxetable}

\section{Discussion and Summary}
\label{sec:Discussion}

We have searched for calcium triplet 
$\lambda \lambda 8500, 8544, 8665$~\AA\ (CaT) absorption in a sample of 17 LRDs in the redshift range $2 < z < 5$ with \emph{JWST} grating data. We detect CaT absorption directly in three sources, and see an average depth of EW$_{\rm CaT} \approx -5$~\AA\ in the sample stack. The detections are clustered in the lowest redshift sources. We argue that LRD infill would make a detection of CaT absorption from stars challenging, meaning the absorption likely arises from the central engine. 

The CaT is a non-resonant transition with a moderate lower-level energy of 1.7 eV from the ground state. CaT in absorption suggests dense warm gas, which can populate the metastable lower level through collisions. Such absorption is seen in stellar photospheres, as well as in some classes of stellar explosions \citep[as summarized in][]{Naidu:2026}. However, to our knowledge, no CaT absorption has ever been detected from the AGN itself, as opposed to CaT from the host (although in the presence of significant host contamination, along with CaT in emission, this would be a challenging observation to make). Thus, CaT absorption becomes another feature of LRDs that sets them apart from classical AGN. This feature can provide new insight into conditions surrounding the central engine. 

The facts above motivate us to associate the CaT absorption with the continuum photosphere of the central engine, but as a caveat we are simply assuming in our model framework that the absorption and continuum radiation emerges at a similar physical location. 
We cannot rule out the alternative scenario where the absorption is instead associated with gas external to the continuum photosphere. The latter is likely the case for Balmer absorbers, which are located no deeper than the broad line region \citep[e.g.,][]{DEugenio:2025,Ji:2025}, but so far we do not have analogous observational evidence for the CaT absorption, and the dissimilar atomic structures between H I and Ca II prevent a meaningful comparison between the two absorbers. Furthermore, we use hydrostatic model atmospheres as a point of comparison, as has been done for the Egg \citep{Liu:2026}, although there is no guarantee that the hydrostatic models can describe the CaT and other LRD properties, nor are they a unique solution to the LRD envelope or cocoon. 

Our model--data comparison suggests that the same model parameters for the Egg predict too much CaT absorption to match the wider sample. However, we find that if the gas is both hotter and less metal-rich in the high redshift systems compared to the Egg, we can reproduce the wider range of EW$_{\rm CaT}$ seen in the sample within the hydrostatic model grid. Alternatively, relaxing the hydrostatic assumption, it could be that the photospheric density falls below the hydrostatic model grid, $\rho_{\rm ph}<10^{-11}{\rm~g~cm^{-3}}$. Then the gas surrounding these LRDs cannot stay in hydrostatic equilibrium due to strong radiative acceleration. 

If weaker CaT is mainly pointing towards higher $T_{\rm eff}$ and less metal-enriched models, then a more comprehensive study could allow insights into the role of metallicity in the LRD phenomenon, particularly since the inferred $\rm[M/H]\approx -2$ seems lower than expectations (albeit indirect) from the mass-metallicity relation. It would be useful to have independent constraints on the metallicity (or metallicity evolution) of LRDs \citep[e.g.,][]{Maiolino:2026,Nikopoulos:2026}. On the other hand, if lower density is the explanation, such densities would aline with other observational evidence for a highly dynamical gas environment around most, if not all, LRDs. The evidence takes the form of detectable asymmetry in broad emission lines \citep{Sneppen:2026}, complicated Balmer absorption line profiles \citep[e.g.,][]{DEugenio:2025,Torralba:2026noon,Chen:2026}, and a finding that the strongest breaks are accompanied by the most systemic -- rather than blue-shifted -- Balmer absorption \citep{Matthee:2026}. 

In the future, simultaneously modeling of the CaT depth and continuum shape should help us distinguish between these possibilities. A joint analysis has been done for the Egg \citep{Liu:2026} and could be extended to high-redshift LRDs given adequate spectral coverage to the rest-IR, where the model prediction differs for the different scenarios.  Long-wavelength data would also constrain the structure of LRDs by probing continuum components in addition to the photosphere, including some dust emission \citep{Lin:2025lowz} or free-free emission \citep{Chisholm:2026,Naidu:2026}. Adding additional absorption features like Ca~H+K could also be powerful, although as resonant lines these transitions are much more sensitive to non-LTE atomic populations and frequency redistribution in the radiative transfer. Ca~H+K has been detected in a couple of LRDs to date \citep[e.g.,][B.~Wang in preparation]{DEugenio:2025}.

While the emission lines of LRDs have been intensively studied, absorption features like CaT, Ca H+K, or the molecular water feature \citep{Wangwater:2026} have received less attention, but may provide important insight into the physical state of the gas surrounding LRDs. A successful model should reproduce the LRD continuum shape, the emission lines, and the absorber properties as well. Furthermore, radiative transfer models incorporating a range of dynamical scenarios could trace the mass of the central object, as prototyped in \citet{Liu:2026}. The primary uncertainties in translating the observed line-width into an enclosed mass are related to geometry (where the line emerges), broadening mechanisms (whether the intrinsic line width and scattering are important), and kinematics (whether the gas motion is virial). However, all of these terms are also significant sources of uncertainty when converting broad emission line FWHM and luminosity into enclosed mass. It is urgent that we develop new means to estimate $M_{\rm BH}$ for the LRDs, and we suggest that the community continue to invest in deeper high-resolution observations of the CaT, along with new models to determine the density and gas conditions leading to this absorption and ultimately to unveil the central engine.

\section*{Acknowledgement}
We benefit from the following JWST programs: UNCOVER (JWST/GO \#2561; Labbé \& Bezanson); ALT (JWST-GO \#3516; Naidu \& Matthee); MegaScience (JWST-GO \#4111; Suess); RUBIES (JWST-GO \#4233; de Graaff \& Brammer); PRIMER (JWST/GO \#1837; Dunlop). We acknowledge funding from NSF/AAG \#2306950, JWST-GO-02561, JWST-GO-03516, and JWST-GO-04111, provided through a grant from STScI under NASA contract NAS5-03127. 

The Cosmic Dawn Center (DAWN) is funded by the Danish National Research Foundation under grant No. 140.
Support for this work was provided by NASA through the NASA Hubble Fellowship grant \#HST-HF2-51618 awarded by the Space Telescope Science Institute, which is operated by the Association of Universities for Research in Astronomy, Incorporated, under NASA contract NAS5-26555.
REH acknowledges support by the German Aerospace Center (DLR) and the Federal Ministry for Economic Affairs and Energy (BMWi) through program 50OR2403 ‘RUBIES’.

\bibliographystyle{aasjournal}
\bibliography{imbh.bib}

\end{document}